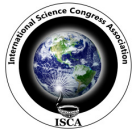

*Review Paper*

# A Method in Security of Wireless Sensor Network based on Optimized Artificial immune system in Multi-Agent Environments

Jaderian Morteza[1], Moradzadeh Hossein[2], Madadipouya Kasra[2], Firoozinia Mohammad[3] and Shamshirband Shahaboddin[4*]
[1]Department of Computer Science, Tabarestan University of Chalous, IRAN
[2]School of Engineering and Technology, Asia Pacific University of Technology and Innovation, MALAYSIA
[3]Institute of Bioscience, University Putra Malaysia, MALAYSIA
[4]Faculty of Computer system and Information Technology, University Of Malaya, MALAYSIA

**Abstract**

*Security in computer networks is one of the most interesting aspects of computer systems. It is typically represented by the initials CIA: confidentiality, integrity, and authentication or availability. Although, many access levels for data protection have been identified in computer networks, the intruders would still find lots of ways to harm sites and systems. The accommodation proceedings and the security supervision in the network systems, especially wireless sensor networks have been changed into a challenging point. One of the newest security algorithms for wireless sensor networks is Artificial Immune System (AIS) algorithm. Human lymphocytes play the main role in recognizing and destroying the unknown elements. In this article, we focus on the inspiration of these defective systems to guarantee the complications security using two algorithms; the first algorithms proposed to distinguish self-nodes from non-self ones by the related factors and the second one is to eliminate the enemy node danger.The results showed a high rate success and good rate of detecting for unknown object; it could present the best nodes with high affinity and fitness to be selected to confront the unknown agents.*

## Introduction

Security in wireless sensor network is important to insure the safety of exchanged information. Lots of techniques have been proposed through Media Access Control (MAC) protocols in computer networks[1] for intelligent selection of sensors[2], headship, and control to increase the data security level. Some algorithm like distributed track finder algorithms (GGWRR2: greedy weighted algorithms)[3] try to find the constant signal field (error) in network through a decoding algorithm including a set of compressed nodes. Computer access protocols (i.e., S-MAC, one of the most important ones) also controls and influences through determining which one of the nodes can access the physical layer (media). The purpose of reducing transfers that occurs at the same time during network access can cause security breach. Artificial immune system algorithms[4] are among of the security methods that draw a lot of attentions by research community. These algorithms inspired by the work of our body's natural immune system tries to recognize unknown agents from self nodes, transmitting signals about intrusion in self nodes, and finally confront these agents to eliminate the danger by ending its power source. Just like natural immune system fighting pathogens, artificial immune systems can be used to protect computer from threats[4,5]. However, natural immune system has other functions such as minimizing the threat to body and making sure that human body system can perform its essential duties normally. The challenge faced by the immune system, when a pathogen inters the body, is to recognize and destroy pathogens (self or non-self) while recognizing self antigens and body molecules[6].

Artificial immune algorithms can be used in broad cases such as pattern recognition and classification[6], and optimization analysis and inquiry. Computer security can be considered as the most important function of artificial immune algorithms[7].

In this paper, we present the review of some existing immune systems and the way of making action when confronting enemy agents in wireless sensor networks. We will also discuss representing necessary and practical algorithms for intrusion detection and confronting the intruder by an exclusive method using the available agents in the networks and representing a software simulation.

## Review of Artificial Immune System

The human immune system is known as very open, un-concentrated and strong system that is able to prepare a high security level against different aggressive organs including bacteria and viruses. Human immune system has the ability to actively recognize and consecutive destroy pathogens. There are firm difficult divisions between body system parts that form innate and adaptive immunity[3]. Natural immune system is known as a multilayer immune system including organic barriers, innate immunity, and adaptive immunity.

Organic Barrier: The first inner layer of human body consists of skin and muscular organ level. Skin provides the first line of defense to invader agents including bacteria, viruses and fungi, by providing both physical and chemical barriers[8].

Innate immunity: Innate immunity (figure 1) or unspecified immunity is the first line of defense against the aggressive agents. It comprises some physical and cellular barriers and mechanisms that provide immediate non-specific defense to the infectious agents.

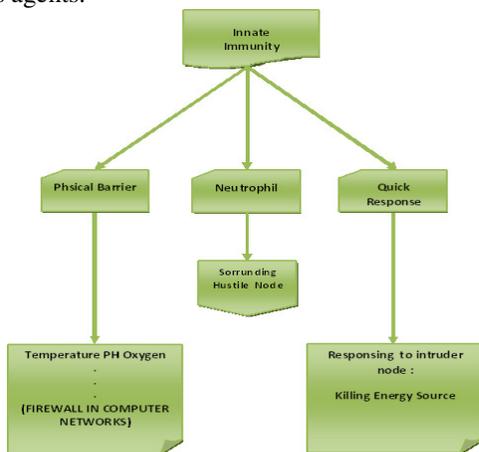

**Figure-1**
**Innate immunity parts in Artificial immine system**

Physical barriers: Epithelial surfaces provide an impermeable physical barrier to a verity of infectious organisms. Peristalsis and cilia eliminate external particles from the respiratory and gastrointestinal tracts; mucus traps bacteria and viruses; adhered bacteria to the skin epithelia can be removed by desquamation.

The phagocyte cells barrier: Some different specific cells like phagocytes, neutrophills and natural killer cells can devour and digest all kind of pathogenic micro-organisms.

Quick response: Active phagocyte tissues produce a variety of signaling compounds named cytokines. These molecules act like hormonal harbingers that send a fiery and firm response to protective organs in the body. i. Adaptive Immunity.

Adaptive immunity, also known as specified or acquired immunity, is a part of immune system composed of specialized immune cells which can clearly recognize and destroy pathogenic microorganisms.

Self and Non-self recognition: The immune system can distinguish between self and non-self cells and response only to non-self molecules.

Immunity memory cells: After recovering from a disease, memory cells are produced from B-cells. Second immune response will be faster and much stronger, when the immune system facing again to the same infectious agent[8].

The consisting elements of human natural body immune system have been shown in (figure 2).

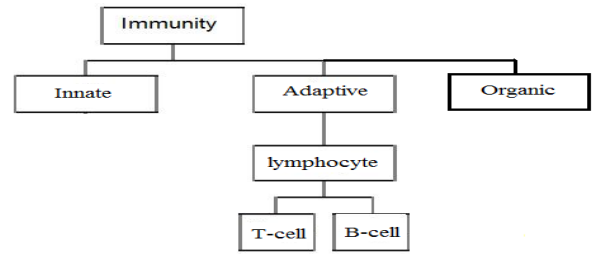

**Figure-2**
**The consistin elements of human immune system**

The immune system is responsible for running a fast response when recognizing a harmful potential pathogen. This process is possible by recognizing some special molecules on the cell surface of pathogens followed by surrounding and destruction of the pathogens[3].

The noticeable point in adaptive immune system would be the white cells or lymphocytes which are divided in two classifications, B-cells and T-cells. Cytotoxic T-cells and T helper cells (Th cells) are two subsets of T-cells. Th cells, also known as $CD^+$ T-cells, can recognize exogenous antigens when they are presented with MHC (Major Histo-compatability complex) class II molecules on the cell surface of a APC (antigen-presenting cells) (figure 3). It has been shown that during a selection process, available T-cells in thymos take the ability of recognition of self cells from non-self. During this process, T helper cells with the affinity of MHC molecules will be placed in positive group; during a negative selection, the cells which are compatible with the self proteins with be eliminated. The remained mature cells have a close affinity to the MHC protein type I and II, however, they have no affinity to the self proteins[2].

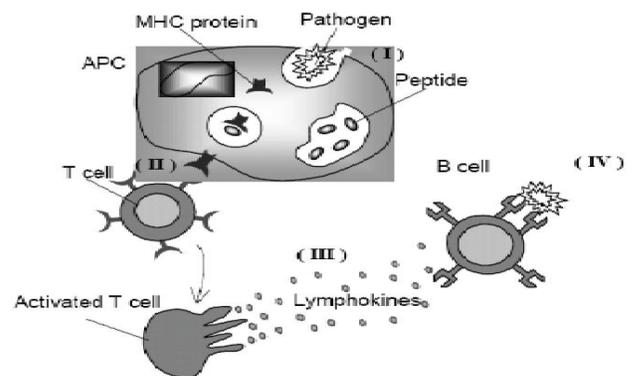

**Figure-3**
**The natural immune system cell devision**

In contrast, B-cells can recognize the pathogens using specific receptors for antigens named antibody. Specific antigen-antibody binding stimulate the B-cells to up take the pathogen during a process called phagocytosis and destroy it by proteolysis enzymes[1]. Together with MHC class II molecules, the process antigens are displayed on the surface of B-cells, where they can be recognized by T helper cells. Once a specific antigen expose to the immune system again, the memory B-cells response result in an immediate and fast reactions without any need to clonal selection or affinity maturity. The details of affinity to antigen are shown in (figure 4).

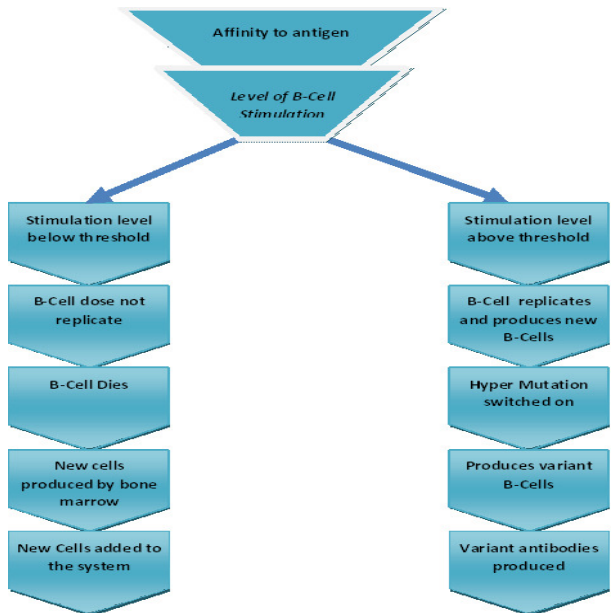

**Figure-4**
**The affinity maturity process chart in antibodies**

Generally, the observed theory and functions from the immunization of artificial immune system have been inspired by the natural immune system principles; these models are used in wide and complicated range of subjects[3]. Although, clonal selection and negative selection are the important part of this system, one of the other algorithms which may have a very high influence on the security of wireless sensor network is an algorithm that first determines the node getting nearer or farther from the network. Thereafter, it determine whether the node is self or non-self and eventually, in case of node being self or non-self, determines the cope kind against the aggressive node. This algorithm is a combination of greedy algorithm and Artificial Immune System (AIS). Negative and clonal selections are accomplished after distinguishing self nodes from non-self nodes to select the best antibodies with the most affinity to the non-self agent. To recognize which nodes are friends and which are enemies, we need to find an algorithm which can recognize chaos and intrusion pattern in non-self (enemy) agents. These nodes can be confronted by adapting suitable techniques, so we will talk briefly about intrusion detection and then represent an algorithm for recognition of the enemy agents.

## Intrusion Detection

Every kind of anomaly (abnormal activity and movements) or incorrect use of computer systems and their source are called intrusion. Intrusion has been divided to the following major categories: i. Computer misuse: forbidden activities by allowed users. ii. Recognition: defining systems or services which may be used in an incorrect way. iii. Intention of intrusion: the forbidden activities to access calculated sources. iv. Intrusion (penetrate): successful access to calculated sources by forbidden users. v. Trojan attacks: the forbidden (unauthorized) process appearance and activities. vi. Denial of access to a service: an attack which result to choke the calculated sources access.

As shown in figure 4, intrusion detection can be seen in two type, anomaly and misuse. Anomaly consists of blocking the network traffic by or toward the host. Hacking changing or erasing the files contents and a system processes results in decreasing the system competency. The misuse is kind of intrusion which intruder tries to use the final user as a back door to upload information and files from the user's computer and also to spread viruses and computer worms. Anomaly detection works by recording the user activities and making profiles from the normal and abnormal behavior of the host network and comparing them to find the intrusion. Whereas, in misuse detection, patterns from MLI (Mark Left by Intrusion) are determined and subsequently the intrusion is recognized by comparing these patterns with the intrusion patterns which were prepared before[9] (figure 5).

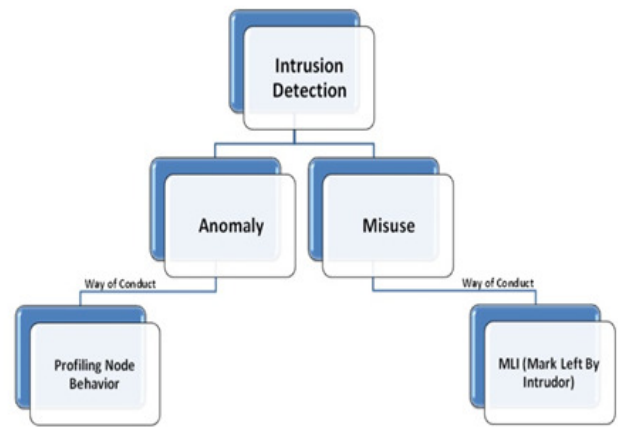

**Figure-5**
**Two different types of intrusion and the confronting method**

Intrusion/anomaly detection has been turned to an important part of computer security by making a new defensive layer against the misuse (abuse) of computer subsequent to the physical layer (like firewall), the identity confirmation layer and access control layer. This layers act in a way that immediately after a node entering our system acceptable limits, the physical layer start working first and closes all the system related entrance and exit for this node. Then, the identity confirmation layer compares the node identity to recognize whether it belongs to this system or it has the membership abilities in the system. In fact, the identity confirmation layer duty is to confirm the identity of related node according to a set of correctness recognition algorithms. If the identity confirmation is done successfully, it means that the considered node can enter the system and acts with the system agents by exerting of a set of access control and limits in access control layer. When the identity confirmation is not successful, and all the node paths to the system will be closed; all the transaction with the system agents will be destroyed; and all of the sources that exchanged in this direction will be thrown away. Anomaly / intrusion detection has the ability to control every node which passed the 3 above layers and entered the system and recognize it, in case of observing abnormal movements or the intention of intrusion in these nodes. Energy in wireless sensor network is very important because the mobile networks are energy-dependant networks[10] and therefore, the anomaly/intrusion detection confronts these nodes by actions like informing other agents and energy reduction of them. The security activities in this layer are done in a way that the security system monitors the entire unique path which are produced the computer user system and programs. It uses a set of different statistical analysis during this supervision to derive the order in this agent's behavioral patterns. This idea originates from the point which the aggressive and invader agent's behavior is considerably different from the legal user's behavior and the detection of this behavior is possible through monitoring these unique tracks.

In contrast with the natural human body system, the danger and threat level is very significant and highly common in computer and computer calculations and this danger level may be increased surprisingly because of some factors like element's destruction and submersible actions[10].

Adavanced algorithm for the detection of node getting nearer or farther and being self or non-self (node filtering): In wireless sensor networks, especially in mobile networks, the nodes getting nearer to our network or their decision to leave the network radiance is very important. Because the exchanged information in WSN can be very crucial, compromising this information can result in changes in battlefield scene (in crucial situations like electronic wars) and changing the winner to a loser or vice -versa. Therefore, presenting an algorithm which can start recognizing the dangerous agents once they enter and then recognizing their types (self or non-self) and finally destroying them from the field may have an effective role in the security of our networks. Since, this algorithm has been inspired by human immune system, knowing the relation between human immune system parts which are directly involved with pathogens and the artificial immune system parts in wireless sensor networks can be useful for a better competence of algorithm, as it shown in table 1. For example, the sent and received string bits in artificial immune system of receivers have been inspired by natural immune system or the lymphocyte T-cells and B-cells are the detector nodes in artificial immune system.

**Table-1**
**Relation between artificial immune system and natural immunr system**

| Immune System | Artificial Immune System |
|---|---|
| Receptor | Bitstring |
| Lymphocyte (B-and T-cell) | Detector |
| Memory Cell | Memory detector |
| Pathogen | Non-self bitstring |
| Binding | Approximate string matching |
| Circulation | Mobile detectors |
| MHC | Representation parameters |
| Cytokines | Sensitivity level |
| Telerization | Distributed negative selection |
| Lymphocyte cloning | Detector replication |

In node filtering algorithm, every node is categorized based on its getting closer or farther and authorized or threat in a wireless sensor network (Algorithm 1). Although this network contains some sensor which each of them prepares just one bit information, a model is used here to determine whether the target is getting closer or farther. In this algorithm, distribution is estimated by random disconnected measurements (see formula 1) defined by nodes and their based weights. An approximation of probability distribution function is derived based on random sampling in the first stage. In the second stage, weights are calculated and normalized for each node based on the math model of formula 1 and in the final stage, a new set of nodes are also called to recognize the unknown agents and to determine whether the node are self or non-self. This node is very important to recognize the nodes which are getting closer or getting farther or suspicious, because if a node which defined as a new node enters the limit of system is suspicious, it will show every behavior except the two predicted situations while identifying by networks agents.

$$\rho(S_i, x_j^{k-1}, x_j^k) = \begin{cases} 1, & if\, s_i^{k-1} \neq s_i^k \\ 1, & if\, s_i^{k-1} = s_i^k\, and\, respect(3) \\ \frac{d(S_i, x_j^k)}{d(S_i, x_j^{k-1})}, & if\, s_i^{k-1} = s_i^k = 1\, and\, threshold < \frac{d(S_i, x_j^k)}{d(S_i, x_j^{k-1})} \leq 1 \\ \frac{d(S_i, x_j^{k-1})}{d(S_i, x_j^k)}, & if\, s_i^{k-1} = s_i^k = -1\, and\, threshold < \frac{d(S_i, x_j^{k-1})}{d(S_i, x_j^k)} \leq 1 \end{cases}$$

**Fourmula-1**
**Calculation probablity distribution function and normalizing the node weight according to the node getting nearer or farther**

Scan all sensors;
Select a specified node;
M, N;
C (a); Closer node
C (b); farther node
While m=1 do
m=m+1
Repeat

For every i in (1, 2….n) = xi € c(a) and xj € c(b);
Check the secure id and party bit (19 bit);
For (j=0;j<19 ;j++)
If (specified id of node && specified parity bit==correct)
// this code checks the validation of id and parity bit
If xi respects network (compatibility)
Then Accept as new node
Send fake packages to new node
If the new node send the item to nodes other than database nodes
Then Accept it as a hustle node else
Accept it as a friend's node
Until number of node==n.

Algorithm 1
The filtering algorithm (identifier of being far or near,self or non-self)

Input: Sseen=set of seen known self elements
Output: D=set of generated detectors
Begin
Repeat
Randomly generate potential detectors and place them in a set P
Determine the affinity of each member of P with each member of the self setSseen
If at least one element in S recognizes a detector in P according to a recognition threshold
Then the detector is rejected, otherwise it is added to the set of available detectors D
Until Stopping Criteria has been met
End.

Algorithm-2
Negative Selection Algorithm

Input: S=set of patterns to be recognized
N=the number of worst element to select for removal
Output =set of memory detectors capable of classifying unseen pattern
Begin
Create an initial random set of antibodies, A for all patterns in S do
Determine the affinity with each antibody in A
Generate clones of a subset of the antibodies in A with the highest affinity
The number of clones for an antibody is proportional to its affinity mutate attributes of these clones to the set A, and place copy of the highest affinity antibodies in A into the memory set ,M
Replace the n lowest affinity antibodies in A with new randomly generated antibodies
End

Algorithm-3
clonal selection algorithm

All the found nodes are first entitled in this security algorithm where a number is specialized to each node. In the next stage, it checks what status the node is in and whether it is getting closer or father to the network (it means if the nodes are entering are getting out of area of scanned nodes). After determining the node status, the system determines whether the node shows a suitable behavior (Good intention) or not from itself, The information sent to the base station from sensor can be another node or access point which comprised of two parts; the first part is sensor code that is able to localize the information sent by the sensor; the second part contains a bit of data which indicates nodes getting closer or farther when sampling a sensor. Thus, it checks for any new enterer node whether it has the valid bits and error code bit (parity bit) or not. If the node is qualified, the next stage of node testing begins (figure 6). Fake packages with un-crucial information will be sent for a new node. Paying attention to this point is necessary that the enemy node has only the duty of collecting and sending data to enemy base station and it doesn't matter what kind of information they are. If this new node sends the new received information to the base, the system would identify this node as enemy and put a suitable way to confront it in its schedule (figure 7). In this situation, using the CRC method to correct errors in parity bit can effectively help in confirming the validity of error code. Adding the validity confirmation stage can result in monitoring and tracking nodes with high accuracy, otherwise the information maybe received inversely to a base computer on our side in an electronic war[11,12].

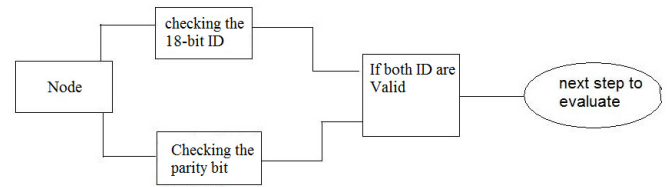

**Figure-6**
**The way of Knowing if the ID and behaviour of nodes are correct**

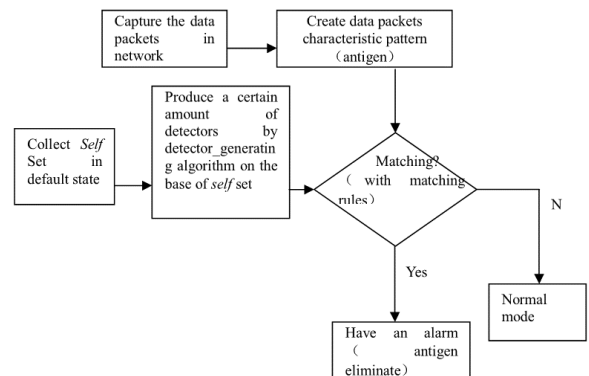

**Figure-7**
**The identifying method followchart**

**Presenting an algorithm to show Negative and Clonal selection antibodies reactions:** First, a brief discussion about negative and clonal selection will be given and then we will engage in description of algorithm function based on this system in wireless sensor networks.

**Negative Selection:** The purpose of negative selection is to make a tolerance for self cells. It is to recognize unknown antigen by the immune system without showing any reaction toward the self cells. During the production of T-cells, receivers are made during a random rearranging process of random genetic. Then, they are exposed to a supervision procedure in thymus called negative selection. The T-cells which reacts toward the self protein would be destroyed, therefore only the cells which do not glue to the self protein are allowed to exit thymus. These mature T-cells patrol in whole body to perform the protective and securitize duties against foreign aggressive anti-gens. The related algorithm to negative selection for WSN is shown in the algorithm 2.

In fact, the negative selection presents another alternative way to recognize patterns. The purpose of pattern recognition is to find the pattern and useful relation from the information in common perspective. This pattern is recognized for saved information about them (patterns) and is considered the difference between these patterns. The complement set of information about the patterns which are saved in special place in this point of view is recognized as available patterns with this related information[8]. The negative selection focuses on anomaly detection problems like intrusion detection in computers and networks (wireless sensor network). As it was discussed before, the purpose of negative selection in natural immune system is self cells tolerance testing or tolerance outset. The T-cells which are in combination with MHC and self amino acids will not be able to finish this test successfully. Virus finding is performed by monitoring the changes in protected and program files by the negative selection algorithm. This algorithm has a lot of advantages in comparison to change monitoring and recognition technique in other methods. The agents that are used in this system are as follow[5,10] (figure 8). i. Supervisor agents, ii. Connector agents, iii. Decided agents which is divided to 3 groups, iv. Helper agents, v. Destroyer agents, vi. Protective agents.

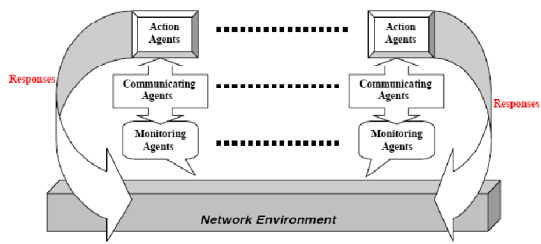

**Figure-8**
**System Agents**

Our system can be located according to the network area situation in the below conditions while the above agents are active: i. Sensing mode: The information sources are monitored in the network area by detection system and are usually used in primary stages to find the abnormal modes. ii. Recognition mode: It makes the suitable decision when a dangerous agent is recognized based on the defensive policies. iii. Response mode:

The detection system put it in performance stage by one of the destroyer, helper or pre-emptive factors after deciding about suitable operations to destroy danger-maker agents[10] (figure 9 and 10).

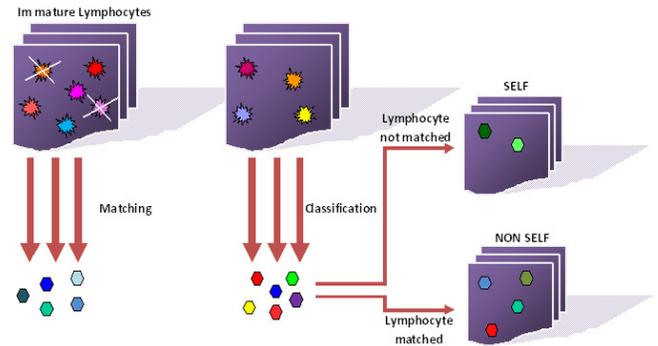

**Figure-9**
**Negative Selection**

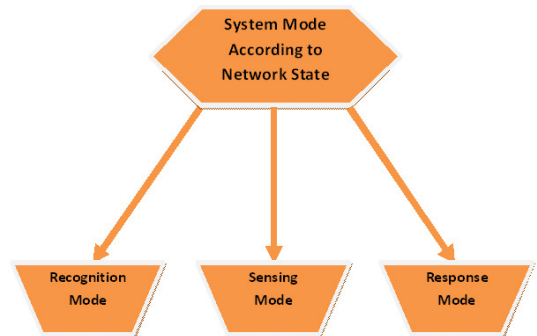

**Figure-10**
**System modes based on network state**

## Clonal Selection Mechanism

In clonal selection, the cells which recognize the anti-gen duplication will be selected. The basic characteristics of clonal selection are: i. The new cells are a copy of their parent's cells (reproduction) which have been exposed to a very high speed maturity mechanism. ii. Destroying the recent changed self-reaction lymphocytes. iii. Changing and duplicating while the mature cells confronting anti-gens.

Antibody-antigen reaction can stimulate B-cells to reproduce its own antibodies. This maturity would be very fast, usually about 1 maturity per 1 cell division. This maturity level lets the immune system to have a fast reaction against antigens. The artificial immune system work method related shown in algorithm 3.

**Responding method to non-self agents Threat:** In artificial immune system, non-self agents such as enemy agents and spy nodes are detected by scanning all the available agents and nodes in the widthwise and linear operation area. This called Local scan because the connection level (affinity) of all the nodes toward an unknown node is measured and the node with the most affinity will be recognized with a put in the head of the confronting team toward the unknown node. This agent (detector) sends messages to close-by nodes to gather around the unknown node. The close nodes will gather around the unknown node once these messages are sent and received by the nearby agents. This action is exactly like cloning (reproduction) in the natural immune system, however, because there is no reproduction is expressed in nodes, the system portraits reproduction play by recalling the nearby node. In confrontation stage against the unknown node, a new innovation was created where ending non-self node energy source technique is used to confront and defeat the node in a way that all the neighbor nodes of the unknown node will send fake packages of information to the unknown node with the purpose of destroying these nodes energy sources. As it was mentioned before, only taking these packages and sending them to enemy base is the duty of these nodes. Gathered agents around the unknown node continue sending fake packages to this node until its energy would be finished; decreasing the node energy to the lowest level shows successful defeating and destruction of the node. In some more complicated systems, there is also a counter-attack after this stage in which a self node misrepresents itself as a non-self node in the system; in other words, it shadows the destructed nodes IP in itself and plays the role of spy node and actually changes the roles in an electronic war (figure 11).

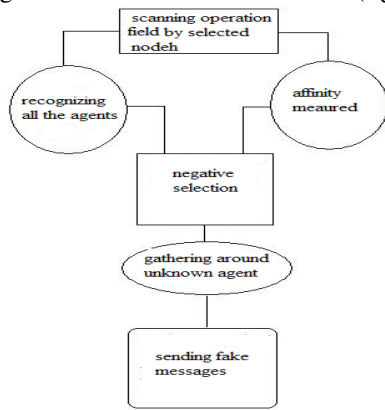

**Figure-11**
**How to confront non- self agents**

The operational examination of negative and clonal selection algorithm in a functional and opertational area: The algorithm which we got familiar with in section 4 of this paper and executed in an operation area and parameters which we used in clonal and negative level are in a way that by using Artificial immune system it will select the most fitness in operational area with the number of antibodies (detectors) of 50, the reproduction (node's recalling) of 20, maturity level (tolerance level) of 80 for antibodies and maximum genes (operational nodes) of 600. The fitness was measured using the following equation:

$$Fitness = (15 \times x \times y \times (1-x) \times (1-y) \times \sin(9 \times \pi \times x) \times \sin(9 \times \pi \times y))^2$$

Where x and y represent the scanned nodes widthwise and linear location in an operational area (two dimensional), respectively. A fitness of 0.8789 can be obtained by the above presupposition information; however, a more fitting result can be reached by changing parameters like antibodies, maximum genes (figure 12).

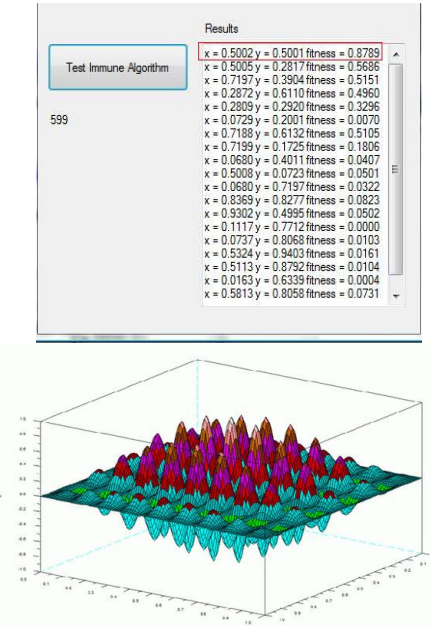

**Figure-12**
**The resulted function in provided software operational area**

The meaning of X and Y in our operational area is the widthwise and linear coordinates of found nodes by detector nodes in a scanned operational area where node fitness was achieved based on their affinity. A three dimensional chart of provided software running result shows the found nodes with the highest fitness in the highest peak of the chart. This figure was provided based on presented presupposition information in the mathematical software. Red peaks in the highest level show best fitness for our experiment and the blue peaks in the lowest level show the worst fitness.

## Conclusion

To confront against aggressive agents, the traditional immune systems have a lot of problems such as the time duration between recognition and action, the lack of recognition and action toward unknown objects until presenting an unresonable behaviour, lack of learning and the lack of relation being in contrast, modern systems, especially the immune systems based on the artificial immune system we discussed in this article do not have these problems. Furthermore, because it has been inspired by the human body immunity system and also because of the learning ability existence, it confronts very effectively against aggressive agents. This system was shown to have a very high rate success and good rate of detection for unknown objects; in this system, could present the best nodes with high affinity and fitness to be selected to confront the unknown agents.